# Asymmetric left-right model and the top pair forward-backward asymmetry

Vernon Barger,[1,*] Wai-Yee Keung,[2,†] and Chiu-Tien Yu[1,‡]

[1]*Department of Physics, University of Wisconsin, 1150 University Avenue, Madison, Wisconsin 53706 USA*
[2]*Department of Physics, University of Illinois, Chicago, Illinois 60607-7059 USA*

The forward-backward asymmetry measurement in top-pair production at the Tevatron is about $2\sigma$ from the standard model prediction. We propose an asymmetric left-right model, which includes a $W'$ boson with a right-handed coupling of down to top quark, and a $Z'$ boson with diagonal couplings to the up, top, and down quarks with $M_{W'} \approx 175$ GeV and $M_{Z'} \approx 900$ GeV. The model accounts for the asymmetry while remaining consistent with the top-pair total cross section and invariant mass distribution.

## I. INTRODUCTION

The top quark is the only fermion whose mass is close to the scale of electroweak symmetry breaking. As such, the top quark is of particular interest in particle physics for it provides a window to many new physics models. Recently, both the CDF and D0 experiments at the Tevatron have observed a sizeable forward-backward asymmetry in top anti-top pair events in which one top decays semileptonically. The recent CDF measurement of the top asymmetry in the $p\bar{p}$ frame, based on 3.2 fb$^{-1}$ of data, is $A_{\text{FB}}^{p\bar{p}} = 0.19 \pm 0.07_{\text{stat}} \pm 0.02_{\text{syst}}$ [1], while the next-to-leading-order (NLO) standard model (SM) prediction [2–5] is $A_{\text{FB}}^{p\bar{p}(\text{SM})} \simeq 0.080$, with a factorization = renormalization scale uncertainty 0.007 [2], but the true theoretical uncertainty may be larger. The discrepancy between the measurement and prediction of the $A_{\text{FB}}^{p\bar{p}}$ is a possible indication of new physics.

Several models have been proposed to explain the $A_{\text{FB}}^{p\bar{p}}$ anomaly. The models are subject to three constraints: $\sigma(t\bar{t})$, $d\sigma/dM_{t\bar{t}}$, and $A_{\text{FB}}^{p\bar{p}}$. The measured $t\bar{t}$ cross section by CDF of $\sigma(t\bar{t}) = 8.1^{+0.98}_{-0.87}$ pb for $m_t = 173.1$ GeV [6] is in good agreement with the SM prediction of $\sigma(t\bar{t})^{\text{SM}} = 7.4 \pm 0.57$ pb for $m_t = 173$ GeV as recently calculated by [7,8] at the next-next-leading-order (NNLO) and by others [9,10] in earlier studies. The invariant mass $M_{t\bar{t}}$ distribution is also in reasonable accord with the SM predictions. In addition to creating the appropriate $A_{\text{FB}}^{p\bar{p}}$, any new physics model must be consistent with the cross section and $t\bar{t}$ mass distribution.

The models proposed to explain the large asymmetry can be placed into two categories. The first consists of models involving the $s$-channel exchange of new vector bosons with chiral couplings to the light quarks and the top quark. The most basic requirement for such a model is a spin-one, color-octet particle with nonzero axial-couplings. Reference [11] provides limits on the axigluon mass while the authors of Ref. [12] consider a chiral color model that involves an axigluon based on the gauge group $SU(3)_A \times SU(3)_B \times SU(2)_L \times U(1)_Y$. The next category consists of models involving the $t$-channel exchange of particles with large flavor-violating couplings. Within this category, the models can be differentiated by the exchange of either a scalar particle $\phi$ or a vector boson. The various possibilities for a scalar particle are limited by the SM gauge structure, and therefore can be categorized by the $SU(3)_C$ representation of $\phi$. The authors of Ref. [13] propose the introduction of a color-sextet or a color-triplet scalar as an explanation of the top quark forward-backward asymmetry. The authors of Ref. [14] propose a $Z'$ boson associated with the $U(1)_{Z'}$ Abelian gauge symmetry with a flavor off-diagonal coupling between the up and top quarks. K. Cheung *et al.* [15] consider a $W'$ boson with off-diagonal right-handed coupling between the down and top quarks. Other recent attempts to address the production asymmetry are found in Refs. [16].

In this article, we propose a model that is based on the gauge group $U'(1) \times SU(2) \times SU'(2)$ with couplings $g'_1$, $g'_2$, and $g'$ associated with the fields $B'$, $W$, $W'$ respectively. We introduce the model in Sec. II and lay out the relations between the couplings. In Sec. III, we estimate the model parameters and discuss potential signatures of our model at the LHC.

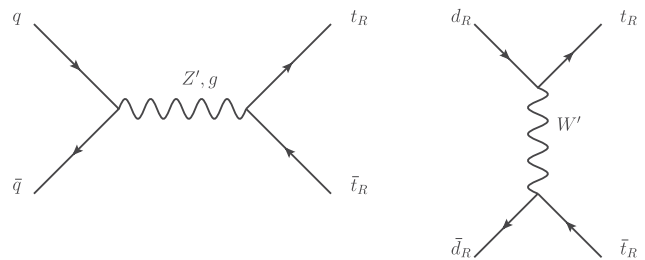

FIG. 1. Tree-level ALRM contributions to $\bar{t}t$ production. $q = d_{L,R}, u_L$ and $\bar{q} = \bar{d}_{L,R}, \bar{u}_L$. The QCD gluon-gluon fusion diagram is not shown; it is included in our calculations but it is subdominant at the Tevatron energy.

*barger@pheno.physics.wisc.edu
†keung@fnal.gov
‡cyu27@wisc.edu

## II. ASYMMETRIC LEFT-RIGHT MODEL

We begin with the gauge group $U'(1) \times SU'(2) \times SU(2)$. The unprimed $SU(2)$ acts on the usual SM left-handed quark doublets. The primed $SU'(2)$ applies to the right-handed doublet $(t, d)_R^T$ in an unconventional grouping. Therefore, the model is a kind of asymmetric left-right model (ALRM) in which we allow different strengths of $L$ and $R$ gauge couplings.

The gauge symmetries are broken sequentially, starting with $U'(1) \times SU'(2) \to U_Y$ to obtain the SM hypercharge $\frac{Y}{2} = T'_3 + \frac{Y'}{2}$, and then $U_Y(1) \times SU(2) \to U_{EM}$ to obtain $Q = T_3 + \frac{Y}{2}$. By using this sequential approximation to the breaking, we preserve the SM interaction.

After symmetry breaking, there are massive $Z'$ and $W'^{\pm}$ bosons in addition to the usual weak bosons. The $W'^{\pm}$ have a $Z'W'W'$ trigauge boson coupling given by $g'^2_2/\sqrt{g'^2_2 + g'^2_1}$, which is of order $\mathcal{O}(1)$. In order to preserve unitarity, the SM $Z$ also appears in the vertex $ZW'^+W'^-$ with coupling $-e\tan\theta_W$. Additional massive fermions will be needed for anomaly cancellation.

Our results are derived by making two successive rotations of gauge boson states. First, we make the rotation

$$B = (g'_2 B' + g'_1 W'^3)/\sqrt{g'^2_1 + g'^2_2},$$
$$Z' = (-g'_1 B' + g'_2 W'^3)/\sqrt{g'^2_1 + g'^2_2}. \quad (1)$$

Then the couplings to the generators become

$$g'_1 \frac{Y'}{2} B' + g'_2 T'_3 W'_3 = \left(\frac{g'_1 g'_2}{\sqrt{g'^2_1 + g'^2_2}}\right) \frac{Y}{2} B + \sqrt{g'^2_1 + g'^2_2}$$
$$\times \left(T'_3 - \frac{g'^2_1 \frac{Y}{2}}{g'^2_1 + g'^2_2}\right) Z'. \quad (2)$$

Subsequently, we perform the usual SM rotation from the basis of $B, W^3$ to the basis of $A, Z$. To simplify the expressions, we denote $g' = \sqrt{g'^2_2 + g'^2_1}$. The SM hypercharge coupling is $g_1 = g'_1 g'_2/g' = e/c_W$, and the SM $SU(2)$ coupling is $g_2 = e/s_W$. It is also useful to note that $g'_1 = (c^2_W/e^2 - g'^{-2}_2)^{-(1/2)}$.

We summarize the neutral current couplings in Table I. We use the usual notation $s_W = \sin\theta_W$ and $c_W = \cos\theta_W$ with $\theta_W$ the weak mixing angle. The second row in the table outlines the generic couplings that apply to various particles as listed in the subsequent rows. Note that $T'_3 = 0$ for the SM left-handed doublets.

The spin- and color-summed amplitude squares for $q\bar{q} \to t\bar{t}$ (Fig. 1) are given by

$$\sum |\mathcal{M}|^2(d_R \bar{d}_R \to t\bar{t}) = (\hat{u} - m^2_t)^2 \left(9 \frac{g'^4_2}{(\hat{t} - M^2_{W'})^2} + 36 \frac{(g'^d_R g'^t_R)^2}{(\hat{s} - M^2_{Z'})^2} + 12 \frac{g'^2_2 g'^d_R g'^t_R}{(\hat{t} - M^2_{W'})(\hat{s} - M^2_{Z'})}\right)$$
$$+ 9g'^4_2 \frac{m^4_t}{M^4_{W'}} \frac{(\hat{t} - m^2_t)^2 + 4M^2_{W'}\hat{s}}{4(\hat{t} - M^2_{W'})^2} + \frac{6g'^2_2 g'^d_R g'^t_R \hat{s}m^4_t/M^2_{W'}}{(\hat{t} - M^2_{W'})(\hat{s} - M^2_{Z'})} + 36(g'^d_R)^2 g'^t_L \frac{g'^t_L(\hat{t} - m^2_t)^2 + 2g'^t_R \hat{s}m^2_t}{(\hat{s} - M^2_{Z'})^2}$$
$$+ 6g'^2_2 g'^d_R g'^t_L \frac{2\hat{s}m^2_t + (\hat{t} - m^2_t)^2 \frac{m^2_t}{M^2_{W'}}}{(\hat{t} - M^2_{W'})(\hat{s} - M^2_{Z'})} + \frac{8g'^2_2 g^2_s/\hat{s}}{\hat{t} - M^2_{W'}}\left[2(\hat{u} - m^2_t)^2 + 2\hat{s}m^2_t + \frac{m^2_t}{M^2_{W'}}[(\hat{t} - m^2_t)^2 + \hat{s}m^2_t]\right]$$
$$+ \frac{16g^4_s}{\hat{s}^2}[(\hat{u} - m^2_t)^2 + (\hat{t} - m^2_t)^2 + 2\hat{s}m^2_t] \quad (3)$$

$$\sum |\mathcal{M}|^2(d_L \bar{d}_L \to t\bar{t}) = \frac{36(g'^d_L)^2}{(\hat{s} - M^2_{Z'})^2}[(g'^t_L)^2(\hat{u} - m^2_t)^2 + (g'^t_R)^2(\hat{t} - m^2_t)^2 + 2g'^t_L g'^t_R \hat{s}m^2_t]$$
$$+ \frac{16g^4_s}{\hat{s}^2}[(\hat{u} - m^2_t)^2 + (\hat{t} - m^2_t)^2 + 2\hat{s}m^2_t] \quad (4)$$

TABLE I. Couplings of the photon, the $Z$-boson, and the $Z'$-boson in the asymmetric left-right model. The second row in the table outlines the generic couplings which applies to various particles as listed in the subsequent rows. We define the coupling strength $g_Z = e/(s_W c_W)$. Note that $T'_3 = 0$ for the SM left-handed doublets.

| neutral boson $f\bar{f}$ pair coupling | $\gamma$ $eQ_q$ | $Z$ $g_Z(T^{SM}_{3,q} - Q_q s^2_W)$ | $Z'$ $g'T'_3 - \frac{g'^2_1}{g'}(Q_q - T^{SM}_{3,q})$ |
|---|---|---|---|
| $u_L \bar{u}_L$ | $\frac{2}{3}e$ | $g_Z(\frac{1}{2} - \frac{2}{3}s^2_W)$ | $\frac{-g'^2_1}{g'}\frac{1}{6}$ |
| $u_R \bar{u}_R$ | $\frac{2}{3}e$ | $-g_Z(\frac{2}{3}s^2_W)$ | $\frac{-g'^2_1}{g'}\frac{2}{3}$ |
| $d_L \bar{d}_L$ | $-\frac{1}{3}e$ | $g_Z(-\frac{1}{2} + \frac{1}{3}s^2_W)$ | $\frac{-g'^2_1}{g'}\frac{1}{6}$ |
| $d_R \bar{d}_R$ | $-\frac{1}{3}e$ | $g_Z(\frac{1}{3}s^2_W)$ | $-\frac{1}{2}g' + \frac{g'^2_1}{g'}\frac{1}{3}$ |
| $t_L \bar{t}_L$ | $\frac{2}{3}e$ | $g_Z(\frac{1}{2} - \frac{2}{3}s^2_W)$ | $\frac{-g'^2_1}{g'}\frac{1}{6}$ |
| $t_R \bar{t}_R$ | $\frac{2}{3}e$ | $-g_Z(\frac{2}{3}s^2_W)$ | $+\frac{1}{2}g' - \frac{g'^2_1}{g'}\frac{2}{3}$ |
| $l_L \bar{l}_L$ | $-e$ | $-g_Z(\frac{1}{2} - s^2_W)$ | $\frac{g'^2_1}{g'}\frac{1}{2}$ |
| $l_R \bar{l}_R$ | $-e$ | $g_Z(s^2_W)$ | $\frac{g'^2_1}{g'}$ |
| $\nu_L \bar{\nu}_L$ | $0$ | $g_Z(\frac{1}{2})$ | $\frac{g'^2_1}{g'}\frac{1}{2}$ |
| $W^+ W^-$, (SM) | $e$ | $e\cot\theta_W$ | $0$ |
| $W'^+ W'^-$ | $e$ | $-e\tan\theta_W$ | $g'^2_2/g'$ |

The $g_{L,R}^{ld}$ and $g_{L,R}^{lt}$ couplings are from Table I, where $g_{L,R}^{ld}$ is the coupling $d_{L,R}\bar{d}_{L,R}Z'$ and $g_{L,R}^{lt}$ is the coupling $t_{L,R}\bar{t}_{L,R}Z'$. Other channels can be obtained by substitutions. The $s$-channel amplitude via $Z'$ interferes with the $t$-channel $W'$, but not with the $s$-channel virtual gluon. The gluon fusion amplitude is the same as in the standard model and can be found in Ref. [17].

It is important to verify that

$$\sum_{V^0=\gamma,Z,Z'} (\text{coupling of } u_L\bar{u}_L \text{ to } V^0)$$
$$\times (\text{coupling of } V^0 \text{ to } W'^+W'^-) = 0. \quad (5)$$

More explicitly,

$$\frac{2}{3}e^2 - \frac{e^2}{c_W^2}\left(\frac{1}{2} - \frac{2}{3}s_W^2\right) - \frac{g_2'^2 g_1'^2}{g'^2}\frac{1}{6} = 0. \quad (6)$$

This guarantees acceptable high energy behavior for the subprocess $u_L\bar{u}_L \to W'^+W'^-$. Incorporating the propagators of $\gamma$, $Z$, $Z'$, we obtain the matrix element squared

$$\sum |M|^2(\bar{u}_L u_L \to W'^+W'^-) = \left[\frac{2}{3}e^2 - \frac{e^2}{c_W^2}\left(\frac{1}{2} - \frac{2}{3}s_W^2\right)\right.$$
$$\left.\times \frac{\hat{s}}{\hat{s}-M_Z^2} - \frac{e^2}{6c_W^2}\frac{\hat{s}}{\hat{s}-M_{Z'}^2}\right]^2$$
$$\times 4A'(\hat{s},\hat{t},\hat{u}). \quad (7)$$

where the $s$-channel function $A'$ is given in Eq. (12) below. In the $s \to \infty$ limit, $A'(s,t,u) \to s/M_{W'}^2$, but the cancellation of couplings renders an acceptable high energy behavior. Similarly, the matrix element squared for $\bar{d}_L d_L \to W'^+W'^-$ is given by

$$\sum |M|^2(\bar{d}_L d_L \to W'^+W'^-) = \left[-\frac{1}{3}e^2 - \frac{e^2}{c_W^2}\left(-\frac{1}{2}+\frac{1}{3}s_W^2\right)\right.$$
$$\left.\times \frac{\hat{s}}{\hat{s}-M_Z^2} - \frac{e^2}{6c_W^2}\frac{\hat{s}}{\hat{s}-M_{Z'}^2}\right]^2$$
$$\times 4A'(\hat{s},\hat{t},\hat{u}) \quad (8)$$

The differential cross sections are given by

$$\frac{d\sigma}{d\hat{t}} = \underbrace{\left(\frac{1}{3}\right)}_{\text{color}}\underbrace{\left(\frac{1}{8\hat{s}}\right)}_{\text{spin-flux}}\underbrace{\left(\frac{1}{8\pi\hat{s}}\right)}_{\text{phase-space}}\sum |M|^2 \quad (9)$$

The charged current interaction

$$\mathcal{L} \supset (g_2'/\sqrt{2})\bar{t}_R\gamma^\mu d_R W'_\mu + \text{H.c.} \quad (10)$$

enters the subprocess $\bar{d}_R d_R \to W'^+W'^-$. The exchange of a right-handed top in the $t$-channel for $\bar{d}_R d_R \to W'^+W'^-$ gives the necessary unitarity cancellation in the high energy limit. The matrix element squared is

$$\sum |M|^2(\bar{d}_R d_R \to W'^+W'^-) = \left(\frac{g_2'}{\sqrt{2}}\right)^4 4E'(\hat{s},\hat{t},\hat{u}) + \left[-\frac{1}{3}e^2 - \frac{s_W^2}{3c_W^2}e^2\frac{\hat{s}}{\hat{s}-M_Z^2} + \left(-\frac{g_2'^2}{2}+\frac{e^2}{3c_W^2}\right)\frac{\hat{s}}{\hat{s}-M_{Z'}^2}\right]^2 4A'(\hat{s},\hat{t},\hat{u})$$
$$+ 2\left(\frac{g_2'}{\sqrt{2}}\right)^2\left[-\frac{1}{3}e^2 - \frac{s_W^2}{3c_W^2}e^2\frac{\hat{s}}{\hat{s}-M_Z^2} + \left(-\frac{g_2'^2}{2}+\frac{e^2}{3c_W^2}\right)\frac{\hat{s}}{\hat{s}-M_{Z'}^2}\right] 4I'(\hat{s},\hat{t},\hat{u}) \quad (11)$$

The functions $A'$, $I'$, $E'$ are

$$A'(\hat{s},\hat{t},\hat{u}) = \frac{1}{4}\left(\frac{\hat{u}\hat{t}}{M_{W'}^4} - 1\right)\left(1 - 4\frac{M_{W'}^2}{\hat{s}} + 12\frac{M_{W'}^4}{\hat{s}^2}\right) + \frac{\hat{s}}{M_{W'}^2} - 4$$

$$I'(\hat{s},\hat{t},\hat{u}) = \left[\frac{1}{4}\left(\frac{\hat{u}\hat{t}}{M_{W'}^4} - 1\right)\left(1 - 2\frac{M_{W'}^2}{\hat{s}} - \frac{4M_{W'}^4}{\hat{s}\hat{t}}\right) + \frac{\hat{s}}{M_{W'}^2} - 2 + 2\frac{M_{W'}^2}{\hat{t}}\right]\frac{\hat{t}}{\hat{t}-m_t^2} \quad (12)$$

$$E'(\hat{s},\hat{t},\hat{u}) = \left[\frac{1}{4}\left(\frac{\hat{u}\hat{t}}{M_{W'}^4} - 1\right)\left(1 + 4\frac{M_{W'}^4}{\hat{t}^2}\right) + \frac{\hat{s}}{M_{W'}^2}\right]\left(\frac{\hat{t}}{\hat{t}-m_t^2}\right)^2$$

The Feynman diagrams for $W'W'$ production are shown in Fig. 2.

Other new vector pairs are $Z'\gamma$, $Z'Z'$, and $WW'$. All three of these processes involve only the $t$-channel. The matrix element for the first two processes can be generically written as

$$\sum |\mathcal{M}|^2(q\bar{q} \to Z'\gamma) = 2(eQ_q)^2[(g_L^{lq})^2 + (g_R^{lq})^2]$$
$$\times \left[\frac{\hat{s}^2 + M_{Z'}^4}{2\hat{t}\hat{u}} - 1\right] \quad (13)$$

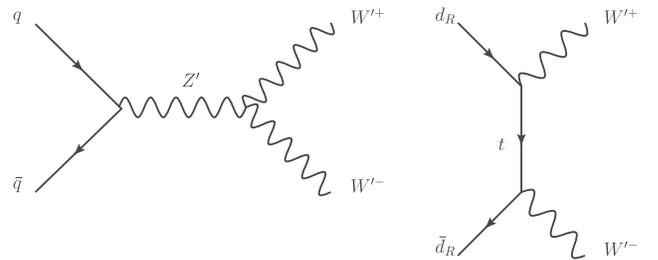

FIG. 2. Tree-level diagrams for the $W'W'$ production in ALRM. $q = d_{L,R}, u_L$ and $\bar{q} = \bar{d}_{L,R}, \bar{u}_L$.

and

$$\sum |M|^2(\bar{q}q \to Z'Z') = [(g_L^{lq})^4 + (g_R^{lq})^4]\left[\frac{\hat{t}}{\hat{u}} + \frac{\hat{u}}{\hat{t}} + \frac{4M_{Z'}^2\hat{s}}{\hat{u}\,\hat{t}}\right.$$
$$\left. - M_{Z'}^4\left(\frac{1}{\hat{t}^2} + \frac{1}{\hat{u}^2}\right)\right] \quad (14)$$

where $q = u_{L,R}, d_{L,R}$, $g_{L,R}^{lq}$ is the left- or right-handed coupling of the quark pair to the $Z'$ boson, and $eQ_q$ is the coupling of the quark pair to $\gamma$. The analogous SM processes are given in the appendix.

The $WW'$ pair production consists of the $t$-channel exchange of a top quark between the bottom and down quarks, and has a matrix element squared of

$$\sum |M|^2(q_iq_j \to WW') = \left[\left(\frac{e}{\sqrt{2}s_W}V_{tq_i}\right)\left(\frac{g_2'}{\sqrt{2}}\right)\right]^2$$
$$\times 4E'(\hat{s}, \hat{t}, \hat{u}) \quad (15)$$

where $q_i = b, d$, $q_j = d_R, \bar{d}_R$ and $V_{tq_i}$ is the CKM mixing matrix element.

## III. TOP PAIR ASYMMETRY AND COLLIDER SIGNALS

The asymmetry $A_{FB}^{p\bar{p}}$ in the $p\bar{p}$ center-of-mass frame is defined as

$$A_{FB}^{p\bar{p}} = \frac{N(\Delta y > 0) - N(\Delta y < 0)}{N(\Delta y > 0) + N(\Delta y < 0)} \quad (16)$$

where $\Delta y = y_t - y_{\bar{t}}$ is the difference in rapidities of the top and anti-top quark. The axial couplings of the $W'$ will contribute to a parity violation in $p\bar{p} \to t\bar{t}$. In order to assess the impact on top-pair measurements at the Tevatron, we implemented our model into MadGraph/MadEvent 4.4.24 [18], using CTEQ6.6M parton distribution functions [19] with factorization and renormalization scales $\mu_F = \mu_R = m_t$ [20]. We took $m_t = 173.1$ GeV [21,22] and applied a uniform SM $K$-factor of $K = 1.31$ [23] to approximate the higher order QCD corrections for $(\text{NNLO}_{\text{approx}})/(\text{LO})$ as shown in [8]. We computed the total cross section $\sigma(t\bar{t})$ for top-pair production [Fig. 4(a)], $A_{FB}^{p\bar{p}}$ [Fig. 4(b)], and $M_{t\bar{t}}$ distribution (Fig. 5) for varying $M_{W'}$, $M_{Z'} = 900$ GeV, $\Gamma_{Z'} = 10.5$ GeV, and $g_2' = 1$. Both $\sigma(t\bar{t})$ and $A_{FB}^{p\bar{p}}$ increase with larger $g_2'$ (Table III) [24].

After accounting for the $\alpha_s^3$ SM contribution to the asymmetry, we are looking for a new physics asymmetry of $A_{FB}^{p\bar{p}}(NP) + 0.080 = 0.19 \pm 0.07$ while reproducing the total cross section $\sigma(t\bar{t}) = 8.1 \pm 0.93$ pb [6]. A comparison of Figs. 4(a) and 4(b) shows that $M_{Z'} \simeq 900$ GeV is compatible with the measured cross section and $A_{FB}^{p\bar{p}}(NP)$ values; the ALRM results of $\sigma(t\bar{t}) = 7.7$ pb and $A_{FB}^{p\bar{p}}(NP) = 0.09$ fall within $1\sigma$ of experimental values. Figure 5 shows the invariant mass distribution for $M_{Z'} = 900$ GeV. Table II lists the reduced chi-square values for

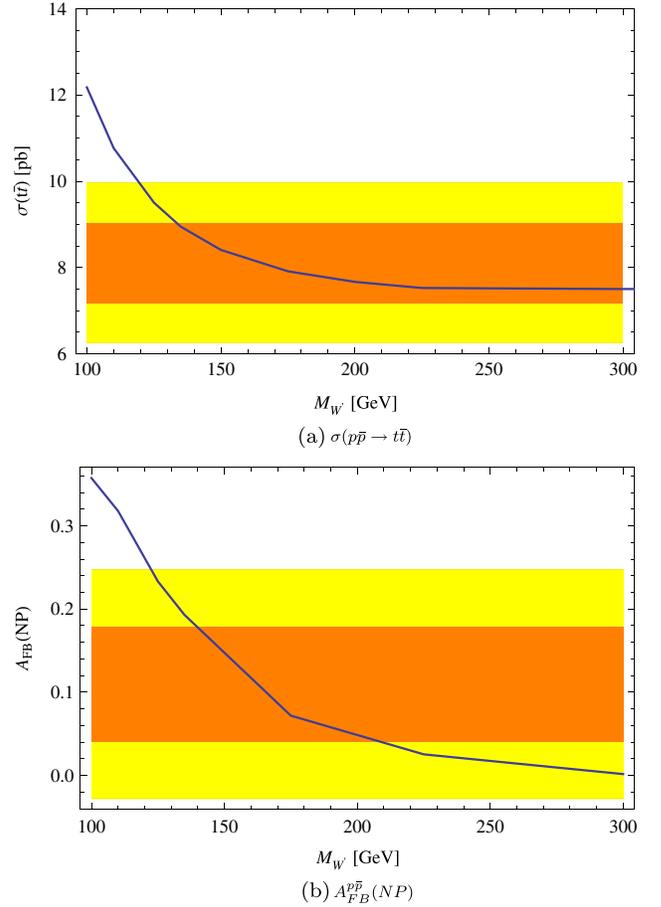

FIG. 4 (color online). (a) $\sigma(t\bar{t})$ [pb] and (b) $A_{FB}^{p\bar{p}}(NP)$ versus $M_{W'}$ in the ALRM with $1\sigma$ (orange) and $2\sigma$ (yellow) CDF bounds.

the various measurements; since the errors are correlated, we do not make best fits to the combined data. Because of the weighting by the parton distribution functions, the $u\bar{u}$ fusion contributions to the cross section dominant over the $d\bar{d}$ fusion contributions by a factor of about 5. However,

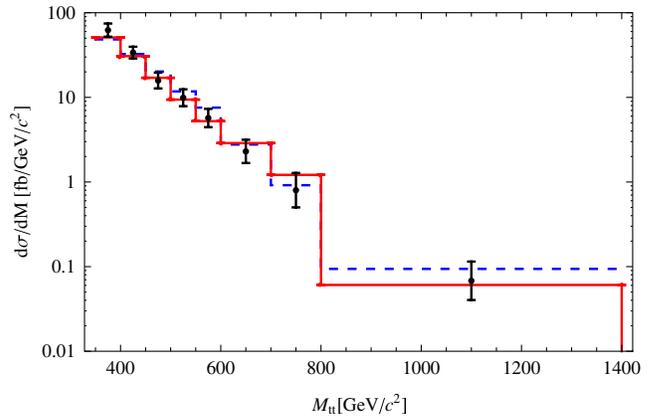

FIG. 5 (color online). $M_{t\bar{t}}$ distribution of CDF data [25], MadEvent SM (red solid line), and ALRM (blue dashed line). $K = 1.31$ has been included uniformly across all bins.

TABLE II. Reduced $\chi^2_{\rm red}$ values for the CDF data versus predictions from the ALRM and the SM.

|  | $M_{Z'} = 900$ GeV | SM |
|---|---|---|
| $A^{p\bar{p}}_{\rm FB}$ | 0.09 | 4.08 |
| $\sigma(t\bar{t})$ | 0.20 | 0.004 |
| $d\sigma/dM_{t\bar{t}}$ | 0.84 | 0.44 |

TABLE III. Comparison of the effects of the value of $g'_2$.

|  | $g'_2 = 0$ | $g'_2 = 0.5$ | $g'_2 = 1$ | $g'_2 = 1.5$ | $g'_2 = 2$ |
|---|---|---|---|---|---|
| $A^{p\bar{p}}_{\rm FB}$ | 0.08 | 0.09 | 0.17 | 0.48 | 0.69 |
| $\sigma(t\bar{t})$ [pb] | 8.10 | 7.49 | 7.68 | 11.6 | 25.2 |

the $W'$ exchange in $d\bar{d}$ fusion is the dominant source of the forward-backward asymmetry (Table IV). The predicted $d\sigma/d\Delta y$ distribution of the ALRM is shown in Fig. 3. The new physics contribution to the cross section comes mostly from $d_R \bar{d}_R$.

The cross sections for $W'$ pair production and $Z' \gamma$ production at the Tevatron are small compared to that of

TABLE IV. Cross sections [pb] of $q\bar{q} \to t\bar{t}$ for $\Delta y > 0$ and $\Delta y < 0$, where $\Delta y = y_t - y_{\bar{t}}$ in the ALRM with $M_{Z'} = 900$ GeV and $g'_2 = 1$. The subprocess CM energy is fixed at $\sqrt{\hat{s}} = 500$ GeV, $\mu_F = \mu_R = M_t = 173.1$ GeV. The QCD contributions to $A_{\rm FB}$ are not included.

| $q\bar{q}$ | $\sigma$ [pb] | $\sigma(\Delta y > 0)$ | $\sigma(\Delta y < 0)$ | $A_{\rm FB}$ |
|---|---|---|---|---|
| $d_L \bar{d}_L$ | 39.7 | 20.1 | 19.3 | ~0 |
| $d_R \bar{d}_R$ | 70.7 | 67.4 | 3.07 | 0.92 |
| $u_L \bar{u}_L$ | 40.0 | 20.0 | 19.6 | ~0 |
| $u_R \bar{u}_R$ | 39.5 | 19.4 | 19.8 | ~0 |
| $gg$ | 19.5 | 9.63 | 9.75 | 0 |

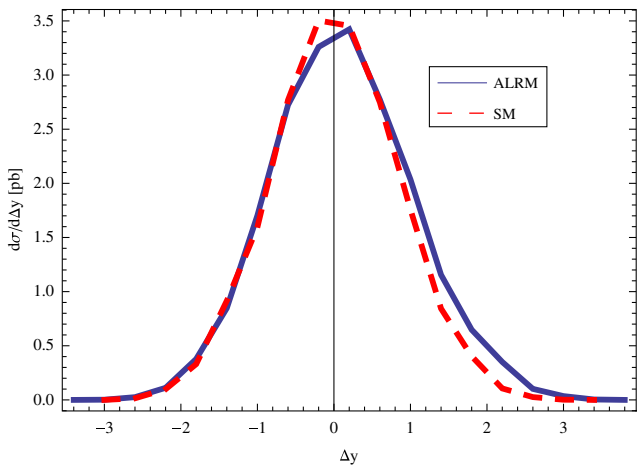

FIG. 3 (color online). Comparison of the $\Delta y = y_t - y_{\bar{t}}$ distribution in the $p\bar{p}$ frame of the ALRM (solid line) with $M_{Z'} = 900$ GeV, $M_{W'} = 175$ GeV, and $g'_2 = 1$ with the SM (dashed line) at the Tevatron.

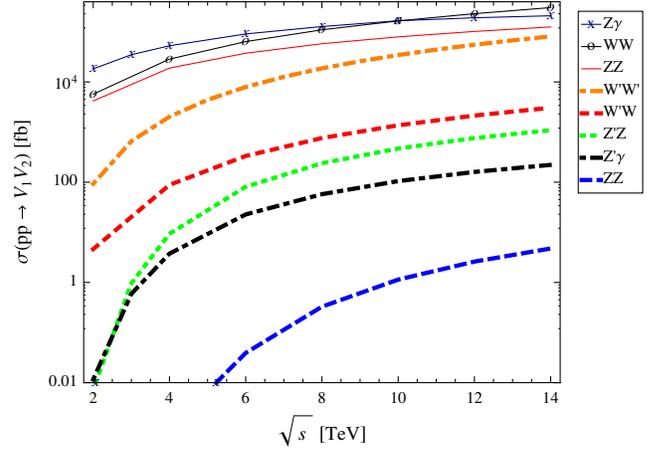

FIG. 6 (color online). Cross-sections [fb] vs $\sqrt{s}$ for various vector boson pairs in $pp$ collisions.

the corresponding SM $W$ processes. We find $\sigma(W'^+ W'^-)/\sigma(W^+ W^-) = 0.03$ and $\sigma(Z' + \gamma)/\sigma(Z + \gamma) = 1.29 \times 10^{-5}$ for $M_{Z'} = 900$ GeV and $g'_2 = 1$.

Figure 6 shows the cross sections for various pairs of vector bosons versus CM energy at the LHC. The SM $WW$, $ZZ$, and $Z\gamma$ cross sections are shown for reference. The expected cross sections at the Tevatron and LHC for these processes are given in Table V.

## IV. $W'$ AND $Z'$ DECAYS

In the limit that the mixings of the new gauge bosons with the SM weak bosons are small, their decays are governed by the interactions in Eq. (9) and Table I. We assume that $m_t < M_{W'} < 200$ GeV. Therefore, the $W'$ decays to a top quark that is almost at rest with respect to the $W'$. For $M_{W'} = 175$ GeV and $g'_2 = 1$, the width for $W'$ decay is

$$\Gamma(W' \to t\bar{d}) = \frac{g'^2_2}{(16\pi)} M_{W'} \left(1 - \frac{3m_t^2}{2M_{W'}^2} + \frac{m_t^6}{2M_{W'}^6}\right)$$
$$= 20.7 \text{ MeV}. \quad (17)$$

This small partial width is due to the limited phase space. For $M_{Z'} = 900$ GeV and $g'_2 = 1$, the partial widths for the leading $Z'$ decays are

$$\Gamma(Z' \to u\bar{u}) = \left(\frac{5g'^2_1}{6g'}\right)^2 \frac{M_{Z'}}{4\pi} = 0.64 \text{ GeV} \quad (18)$$

$$\Gamma(Z' \to d\bar{d}) = \left[\left(\frac{g'^2_1}{6g'}\right)^2 + \left(\frac{g'^2_1}{3g'} - \frac{g'}{2}\right)^2\right] \frac{M_{Z'}}{8\pi} = 8.60 \text{ GeV} \quad (19)$$

$$\Gamma(Z' \to l\bar{l}) = \left(\frac{3g'^2_1}{2g'}\right)^2 \frac{M_{Z'}}{8\pi} = 0.86 \text{ GeV} \quad (20)$$

The $Z'$ can be singly produced in the s-channel at the Tevatron, but the dijet signal from its decays will be difficult to identify above the large QCD dijets back-ground. One can set a lower limit on the $Z'$ mass from dimuon and $t\bar{t}$ production. The $M_{t\bar{t}}$ distribution gives a more stringent limit, $M_{Z'}^{\text{ALRM}} \gtrsim 800$ GeV, than the dimuon data give.

TABLE V. Expected values for $\sigma(V_1 V_2)$ [pb] at the Tevatron ($\sqrt{s} = 2$ TeV) and the LHC ($\sqrt{s} = 14$ TeV). $K = 1.31$ has been included.

|  | $W^+W^-$ | $W'^+W'^-$ | $Z'\gamma$ | $Z'Z'$ | $WW'$ | $Z\gamma$ | $ZZ$ | $ZZ'$ |
|---|---|---|---|---|---|---|---|---|
| Tevatron | 15.12 | 0.49 | 0.003 | ~0 | 0.02 | 23.33 | 8.52 | ~0 |
| LHC | 304.30 | 81.14 | 0.17 | 0.005 | 3.06 | 159.67 | 125.16 | 1.09 |

## V. SUMMARY

We have proposed an asymmetric left-right model based on the $U'(1) \times SU'(2) \times SU(2)$ gauge group. The symmetry is broken spontaneously, first by a Higgs doublet of the prime sector to $U_Y$, and then by another Higgs doublet in the SM sector. The $SU'(2)$ acts on a $(t, d)_R$ doublet. The ALRM includes a $W'$ boson with the $(t, d)_R$ coupling, and a $Z'$ boson with diagonal $u\bar{u}$, $d\bar{d}$, and $t\bar{t}$ couplings. With $M_{W'} \approx 175$ GeV and $M_{Z'} \approx 900$ GeV, the ALRM can explain the $A_{FB}^{p\bar{p}}$ measurement at the Tevatron, while remaining consistent with the $\sigma(t\bar{t})$ and $t\bar{t}$ mass distribution [1,6,25]. We have evaluated the cross sections for the production of vector boson pairs at the LHC. Since the $W'$ decays only to quarks, its collider signal may be difficult to distinguish from SM backgrounds. However, small mixings with the $W$ will lead to small leptonic branching fractions that should allow it to be more readily probed at the LHC. The $Z'$ can be probed via dilepton and $t\bar{t}$ production.

## ACKNOWLEDGMENTS

The authors thank Q.-H. Cao, N. Christensen, Y. Gao, I. Lewis, M. McCaskey, and C. Wagner for helpful discussions. This work was supported in part by the U.S. Department of Energy under Grant Nos. DE-FG02-95ER40896 and DE-FG02-84ER40173. C.-T. Yu is supported by the National Science Foundation.

## APPENDIX

Below are formulas for $W^+W^-$ production in the SM.

$$\sum |M|^2(\bar{d}_L d_L \to W^+W^-)$$

$$= \left(\frac{e/s_W}{\sqrt{2}}\right)^4 4E(\hat{s}, \hat{t}, \hat{u}) + \left[-\frac{1}{3}e^2 + \left(-\frac{1}{2} + \frac{s_W^2}{3}\right)\frac{e^2}{s_W^2}\right.$$
$$\left. \times \frac{\hat{s}}{\hat{s} - M_Z^2}\right]^2 4A(\hat{s}, \hat{t}, \hat{u}) + 2\left(\frac{e/s_W}{\sqrt{2}}\right)^2 \left[-\frac{1}{3}e^2\right.$$
$$\left. + \left(-\frac{1}{2} + \frac{s_W^2}{3}\right)\frac{e^2}{s_W^2}\frac{\hat{s}}{\hat{s} - M_Z^2}\right] 4I(\hat{s}, \hat{t}, \hat{u})$$

and

$$\sum |M|^2(\bar{u}_L u_L \to W^+W^-)$$

$$= \left(\frac{e/s_W}{\sqrt{2}}\right)^4 4E(\hat{s}, \hat{u}, \hat{t}) + \left[\frac{2}{3}e^2 + \left(\frac{1}{2} - \frac{2s_W^2}{3}\right)\frac{e^2}{s_W^2}\right.$$
$$\left. \times \frac{\hat{s}}{\hat{s} - M_Z^2}\right]^2 4A(\hat{s}, \hat{u}, \hat{t}) - 2\left(\frac{e/s_W}{\sqrt{2}}\right)^2$$
$$\times \left[\frac{2}{3}e^2 + \left(\frac{1}{2} - \frac{2s_W^2}{3}\right)\frac{e^2}{s_W^2}\frac{\hat{s}}{\hat{s} - M_Z^2}\right] 4I(\hat{s}, \hat{u}, \hat{t})$$

Note the interchange of $u \leftrightarrow t$ in the functions for the $u_L$ subprocess. The $\bar{u}_R u_R$ and $\bar{d}_R d_R$ processes only involve the $s$-channel in the SM,

$$\sum |M|^2(\bar{d}_R d_R \to W^+W^-) = \left(-\frac{1}{3}e^2\right)^2 \left(\frac{M_Z^2}{\hat{s} - M_Z^2}\right)^2$$
$$\times 4A(\hat{s}, \hat{t}, \hat{u})$$

and

$$\sum |M|^2(\bar{u}_R u_R \to W^+W^-) = \left(\frac{2}{3}e^2\right)^2 \left(\frac{M_Z^2}{\hat{s} - M_Z^2}\right)^2 4A(\hat{s}, \hat{u}, \hat{t})$$

The formulas for the other vector boson pairs are [27]:

$$\sum |\mathcal{M}|^2(q\bar{q} \to Z\gamma) = 2(eQ_q)^2 (g_L^{q2} + g_R^{q2})\left[\frac{\hat{s}^2 + M_Z^4}{2\hat{t}\hat{u}} - 1\right],$$

$$\sum |M|^2(\bar{q}q \to ZZ) = ((g_L^q)^4 + (g_R^q)^4)\left[\frac{\hat{t}}{\hat{u}} + \frac{\hat{u}}{\hat{t}} + \frac{4M_Z^2\hat{s}}{\hat{u}\hat{t}}\right.$$
$$\left. - M_Z^4\left(\frac{1}{\hat{t}^2} + \frac{1}{\hat{u}^2}\right)\right]$$

and

$$\sum |\mathcal{M}|^2(q_i\bar{q}_j \to ZW^\pm) = \frac{2e^4}{s_W^2}|V_{ij}|^2\Bigg\{\left(\frac{1}{\hat{s}-M_W^2}\right)^2\left[\left(\frac{9-8s_W^2}{4}\right)(\hat{u}\hat{t}-M_W^2 M_Z^2) + (8s_W^2-6)\hat{s}(M_W^2+M_Z^2)\right]$$
$$+\left[\frac{\hat{u}\hat{t}-M_W^2 M_Z^2 - \hat{s}(M_W^2+M_Z^2)}{\hat{s}-M_W^2}\right]\left(\frac{g_L^{q_j}}{\hat{t}}-\frac{g_L^{q_i}}{\hat{u}}\right) + \frac{\hat{u}\hat{t}-M_W^2 M_Z^2}{4(1-s_W^2)}\frac{1}{g_Z^2}\left(\frac{g_L^{q_j 2}}{\hat{t}^2}+\frac{g_L^{q_i 2}}{\hat{u}^2}\right)$$
$$+\frac{\hat{s}(M_W^2+M_Z^2)}{2(1-s_W^2)}\frac{g_L^{q_j} g_L^{q_i}}{g_Z^2 \hat{t}\hat{u}}\Bigg\}$$

where $g_{L,R}^q$ is the SM coupling between the quark pair and the $Z$ boson, and $eQ_q$ is the SM coupling between the quark pair to $\gamma$. The functions $A$, $I$, $E$ of Ref. [28] are

$$A(\hat{s},\hat{t},\hat{u}) = \frac{1}{4}\left(\frac{\hat{u}\hat{t}}{M_W^4}-1\right)\left(1-4\frac{M_W^2}{\hat{s}}+12\frac{M_W^4}{\hat{s}^2}\right) + \frac{\hat{s}}{M_W^2} - 4$$

$$I(\hat{s},\hat{t},\hat{u}) = \left[\frac{1}{4}\left(\frac{\hat{u}\hat{t}}{M_W^4}-1\right)\left(1-2\frac{M_W^2}{\hat{s}}-\frac{4M_W^4}{\hat{s}\hat{t}}\right) + \frac{\hat{s}}{M_W^2} - 2 + 2\frac{M_W^2}{\hat{t}}\right]$$

$$E(\hat{s},\hat{t},\hat{u}) = \left[\frac{1}{4}\left(\frac{\hat{u}\hat{t}}{M_W^4}-1\right)\left(1+4\frac{M_W^4}{\hat{t}^2}\right) + \frac{\hat{s}}{M_W^2}\right]$$

---